\documentclass[prl,twocolumn,showpacs,superscriptaddress]{revtex4}
\usepackage{graphicx,amsmath,amssymb,natbib}

\newcommand{\figwidth}{0.90\columnwidth}

\newcommand{\Fig}[1]{Fig.~\ref{#1}}
\newcommand{\Figlong}[1]{Figure~\ref{#1}}

\newcommand{\kbt}{ k_{\rm B}T }

\begin{document}

\title{Novel crystal phase in suspensions of hard ellipsoids}

\author{P. Pfleiderer}
\email{pfleider@uni-mainz.de}
\author{T. Schilling}
\affiliation{Institut f\"ur Physik, Johannes Gutenberg-Universit\"at,
Staudinger Weg 7, D-55099 Mainz, Germany}

\date{\today}

\begin{abstract} 
We present a computer simulation study on the crystalline phases of hard
ellipsoids of revolution. For aspect ratios $\ge 3$  the previously
suggested stretched-fcc phase [D. Frenkel and   B. M. Mulder,
  Mol. Phys. {\bf 55}, 1171 (1985)] is replaced by a novel
crystalline phase. Its unit cell contains two ellipsoids with
unequal orientations. The lattice is simple monoclinic.
The angle of inclination of the lattice, $\beta$, is a very soft 
degree of freedom, while the two right angles are stiff. 
For one particular value of $\beta$, the close-packed version of this crystal
is a specimen of the family of superdense packings
recently reported [Donev et al., Phys. Rev. Lett. {\bf 92}, 255506 (2004)]. 
These results are relevant for studies of nucleation and glassy dynamics of 
colloidal suspensions of ellipsoids.
\end{abstract}


\maketitle

Classical, hard particles such as non-overlapping spheres, rods
or ellipsoids are widely used as models for granular matter, 
colloidal and molecular fluids, crystals and glasses. Their
success---and their appeal---lies in the fact that the problem of
evaluating a
many-body partition function is reduced to a slightly simpler, geometrical
problem, namely the evaluation of entropic contributions only. This is an
advantage, in particular, for computer simulations. Hence one of the first
applications of computer simulations was a study of the liquid-solid
phase transition in hard spheres \cite{alder.wainwright:1957}.

In this Letter, we re-examine the high-density phase behavior of hard
ellipsoids of revolution with short aspect ratios. This system has
been studied in Monte Carlo simulations by Frenkel and Mulder in 1985
\cite{frenkel.mulder:1985}. Since then, the focus of attention has
been on the nematic phase and the isotropic-nematic transition
\cite{zarragoicoechea.levesque.weis:1992,allen.mason:2002,
camp.mason.allen.khare.kofke:1996}. Biaxial hard ellipsoids have also
been studied \cite{allen:1990,camp.allen:1997}.
But to our knowledge, the high-density phases have not been
investigated further.
Knowledge of these phases is relevant for studies of
elongated colloids in general, and it is crucial for the study of
nucleation \cite{schillingt.frenkel:2005} and glassy
dynamics \cite{letz.schillingr.latz:2000} in hard ellipsoids.

At high densities, Frenkel and Mulder assumed that the most stable phase was 
an orientationally ordered
solid which can be constructed in the following way: A face-centered
cubic (fcc) system of spheres is stretched by a factor
$x$ in an arbitrary direction $\bf{n}$. This transformation results
in a crystal structure of
ellipsoids of aspect ratio $x$, which are oriented along $\bf{n}$. As the
transformation is linear, the density
of closest packing is the same as for the closest packing of spheres 
$\phi = \pi/\sqrt{18} \approx 0.7405$. Recently, Donev and co-workers
showed that ellipsoids can
be packed more efficiently if non-lattice periodic packings (i.e.
packings in which a unit cell contains several ellipsoids at different
orientations) are taken into account \cite{donev.etal:2004}. For
unit cells containing two particles, they constructed a family of
packings which reach a density of $\phi = 0.770732$ for aspect
ratios larger than $\sqrt{3}$.

We have performed Monte Carlo simulations of hard ellipsoids of 
revolution with aspect ratios $a/b = \frac{1}{3},2,3,4,6$ and found 
that for large 
parts of the high density phase diagram, the lattice crystal suggested by 
Frenkel and Mulder is unstable with respect to a different crystalline 
phase. Its unit cell is simple monoclinic and contains two ellipsoids at an
angle with respect to each other. We will refer to this new phase as
SM2 (simple monoclinic with a basis of two ellipsoids).

Simulations were performed at constant particle number $N$, pressure $P$ and
temperature $T$. The shape of the periodic box was allowed to fluctuate,
so that the crystal unit cell could find its equilibrium shape. This was
achieved by implementing the Monte Carlo equivalent of the simulation
method by Parrinello and
Rahman \cite{parrinello.rahman:1981, najafabadi.yip:1983, yashonath.rao:1985}.
We constructed the initial solid structures by
stretching an fcc hard sphere crystal along the [111] direction by a
factor of $a/b$. Hence the simulations were started with
crystals identical to the ones studied by Frenkel and Mulder.
For aspect ratios $a/b = \frac{1}{3},2,3$, we simulated eight
independent systems, each containing $N\approx 1700$ particles. 
Simulations started at $P=50\;\kbt/8ab^{2}$.
The pressure was lowered in
subsequent runs until we observed melting to the nematic
phase. In each run, equilibration lasted roughly 2 million MC
sweeps, and was followed by 1.8 to 3.2 million MC sweeps for calculating
thermodynamic averages. 
(One MC sweep consisted of $N$ attempts to move or rotate a
particle and one attempt to change the box shape, on average.)
For aspect ratio $a/b = 6$, we simulated eight independent systems
with $N=3072$ at a pressure $P=46~\kbt/8ab^{2}$. Equilibration and data
acquisition summed up to more than 3 million MC sweeps.
One system with $a/b=4$, $N=1200$ and $P=46~\kbt/8ab^{2}$ was first
simulated sampling only rectangular box shapes for a total of
2.2 million MC sweeps, and was then simulated including
non-rectangular box shapes
at the same and lower pressures for several million MC sweeps.

Particle overlap was checked by a routine \cite{allen.frenkel.talbot:1989}
which uses the
Vieillard-Baron \cite{vieillardbaron:1972} and
Perram-Wertheim \cite{perram.wertheim:1984, perram.wertheim:1985}
criteria. In a simulation of a system with $a/b = 3$ and $N=1728$, 1
million MC sweeps took about 12 hours of CPU time on a $1.8~{\rm GHz}$
processor.

All systems with aspect ratios $a/b \geq 3$ left the initial fcc
structure in favor of a simple monoclinic lattice with a basis of two
ellipsoids (SM2). We will
discuss this structure for $a/b = 3$ and return to different aspect
ratios towards the end of this Letter.


\begin{figure}
\begin{center}
\includegraphics[ 
  width=\figwidth
  ]{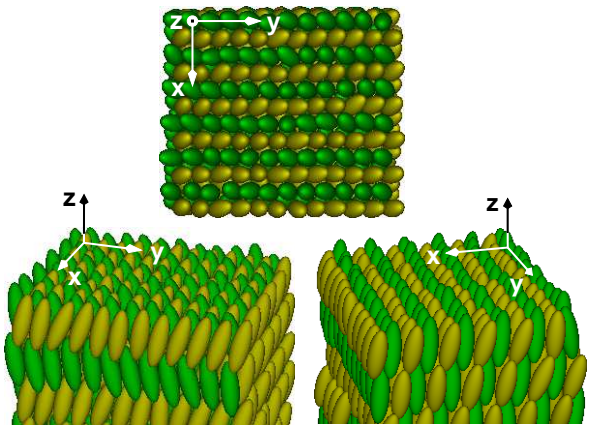}
\caption{\label{snap} Snapshot of the SM2 crystal from different
  angles, with $a/b = 3$, $N=1728$, and $P = 46~\kbt/8ab^{2}$. The online
  version is color coded with respect to orientation.}
\end{center}
\end{figure}

\Figlong{snap} shows a snapshot of a system in the SM2
phase with $a/b = 3$.
The color code (online) helps distinguish the two directions of
orientation which are present in the crystal. In the initial configuration, a
stretched-fcc crystal, 
all ellipsoids were oriented along the $z$-axis. SM2 is mainly the
result of a collective re-orientation. The two directions of orientation
alternate from layer to layer. Layers containing ellipsoids of only one
orientation are here parallel to the $y$-axis and form an angle with the 
$x$-axis. Within layers parallel to the 
$xy$-plane, the centers of mass of the ellipsoids form a nearly
triangular lattice. It differs from the initial fcc crystal by a
slight elongation along the $x$-axis. However, the collective
re-orientation of the ellipsoids displaced their \emph{tips} in such a
way that they now form a rectangular lattice. This can be discerned in
the top view in \Fig{snap}. The tips of the neighboring layers
interlace. As a result, each ellipsoid now has four nearest neighbors
above and below, whereas in fcc, it had three. The total number of
nearest neighbors has increased from 12 to 14, which is indicative of
a higher packing efficiency. 

The initial triangular symmetry about the $z$-axis allows for
two additional, equivalent SM2 configurations, which
are rotated with respect
to the one in \Fig{snap} by $\pm 60^{\circ}$. We observed these
possibilities as well; in fact, only two out of our eight systems
assumed the global orientation seen in \Fig{snap}.


\begin{figure}
\begin{center}
\includegraphics[ 
  width=0.8\columnwidth
  ]{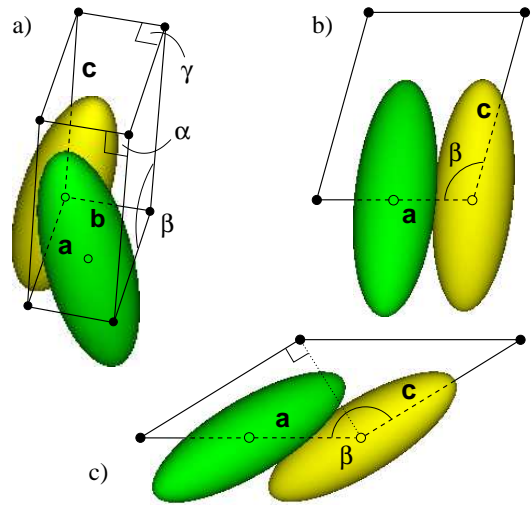}
\caption{\label{unitcell} Unit cell of SM2 with $a/b = 3$ (color
  online). The open circles indicate the centers of the two ellipsoids
  which form the basis. The cell is monoclinic. $\beta$ exhibits large
  variations. The yellow (light gray) ellipsoid is at the origin, the
  green (dark gray) one is at $\frac{1}{2}({\bf a}+{\bf b})$. The
  orientations are symmetric about the $ac$-plane. The parameters in
  Parts a) and b) are average values for $N=1728$ and $P =
  46~\kbt/8ab^{2}$; cf.\ \Fig{snap}. Part c) shows the cell at close
  packing with $\beta = 148.3^\circ$, where it is an instance of the
  family of packings introduced by Donev et al.\ \cite{donev.etal:2004}. }
\end{center}
\end{figure}

The unit cell of SM2 is shown in \Fig{unitcell}. The open circles
indicate the centers of the two ellipsoids which form the basis. The
cell is monoclinic. The yellow (light gray) ellipsoid is at the
origin, the green (dark gray) one is at $\frac{1}{2}({\bf a}+{\bf
  b})$. The orientations are symmetric about the $ac$-plane. The
parameters used to produce parts a) and b) of \Fig{unitcell} are 
thermal average values obtained from simulations with
$N=1728$ and $P = 46~\kbt/8ab^{2}$; cf.\ \Fig{snap}. The cell remained
monoclinic even when the pressure was lowered down to the melting transition 
into the nematic phase. 

The angle of inclination, $\beta$, relaxes extremely slowly. The 
simulations with $N=1728$ were too slow to equilibrate this angle. 
Therefore we carried out a set of very long simulations for a smaller 
system ($N=432$) with initial values of 
$\beta$ in the range $105^\circ < \beta < 150^\circ$. After more than
100 million Monte Carlo sweeps, there was still no clear evidence for a
preferred value of $\beta$. Variations of $15^\circ$ in a single
simulation were typical, even at $P = 46~\kbt/8ab^{2}$. Hence, we expect 
the shear modulus in this degree of
freedom to be very small. The other two angles, $\alpha$ and $\gamma$,
were stable at $90^\circ$, with fluctuations of $< 1^\circ$. The
associated shear moduli are much larger. The reason for this
interesting rheological property is that planes of uniform orientation 
slide well
past each other in the {\bf c}-direction only. In some of the long
simulations, this led to undulations of the lattice in the {\bf
  c}-direction to the point of planar defects, which would
spontaneously heal again. 

To find a lower bound for the maximum density of SM2 ($a/b=3$), we
performed simulations sampling only the unit cell parameters and
particle orientations, and imposing all symmetries of SM2. The initial
parameters were average values obtained from the simulations with
$N=1728$ and $P = 46~\kbt/8ab^{2}$. In the process of maximizing the
density, $\beta$ increased from $105^\circ$ to about $150^\circ$, and
the common tilt of the ellipsoids with respect to the $bc$-plane
disappeared. We then imposed that $({\bf a}+{\bf c})$ be
perpendicular to ${\bf c}$ (see \Fig{unitcell} Part c)) which is equivalent to
$\beta \approx 148^\circ$. Under this condition 
SM2 becomes an instance of the family of packings introduced by Donev et al. 
This simulation achieved the highest packing fraction, namely 
$\phi \approx 0.770732$ (the value reported by Donev et al.).

But already at $\beta \approx 105^\circ$ we found a
jamming density of $99.663\%$ of the maximum. Simulations at intermediate
values indicate a smooth approach towards the maximum density as
$\beta$ increases. Thus, the close-packing density varies very weakly
for $105^\circ < \beta < 148^\circ$. While this range is
traversed, ellipsoids of one orientation move past neighbors of the
other orientation by almost half their length. This can be seen in
Parts b) and c) of \Fig{unitcell}. For reasons of symmetry, this
translation may even continue by the same amount while the density remains
above $99.663\%$ of the maximum. These observations are in accord with
the fact that $\beta$ is soft at finite pressures; evidently, the
free volume distribution possesses a similarly slight variation with $\beta$.

\begin{figure}
\begin{center}
\includegraphics[ 
  width=\figwidth
  ]{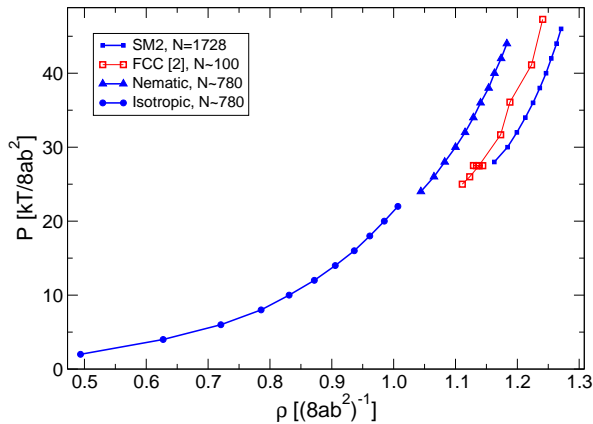}
\caption{\label{eqState} Equation of state data for $a/b=3$ and $N=1728$. 
The open squares show
  stretched-fcc data by Frenkel and Mulder \cite{frenkel.mulder:1985},
  the filled squares the higher-density SM2 phase. Also shown are the
  nematic and fluid branches (triangles and circles,
  respectively). Errors on our data are indicated by the size of the
  symbols. The SM2 curve tends to underestimate the density slightly 
  since $\beta$ was not equilibrated entirely.
}
\end{center}
\end{figure}

\Figlong{eqState} shows equation of state data of SM2 from our simulations
with $N=1728$ particles. The 
density of SM2 is higher than that of stretched fcc for all pressures. Five
of eight systems underwent the transition to SM2 already at the
highest simulated pressure $P = 48~\kbt/8ab^{2}$, the remaining
three at $P = 46~\kbt/8ab^{2}$.  Note also that in all our runs, SM2 melted
to the nematic phase without re-visiting the stretched fcc phase from
which it developed; evidently, SM2 does not only pack more efficiently than
stretched fcc, it also provides for a better distribution of free 
volume at all densities until the transition to the nematic. 
Hence it is more stable than stretched fcc.
We also show the nematic branch from an $(N,P,T)$ compression
(i.e.\ the pressure was raised between successive simulations)
with $N \approx 780$ particles and up to 6 million MC sweeps per
run. Even at strong over-compression, no spontaneous
crystallization occurred. This indicates that the nucleation barrier
to the SM2 
phase is very high. Also shown is the isotropic fluid branch as obtained
from $(N,P,T)$ compression and expansion runs with $N \approx 780$.

All eight simulations at $a/b = 6$ and $P = 46~\kbt/8ab^{2}$
formed SM2 as well, although four of
them retained a planar defect. Different regions
in the periodic box were able to develop different global orientations
of SM2 as the systems were larger ($N = 3072$) than those with $a/b =
3$ ($N = 1728$). 
We also simulated a system with $a/b = 4$, $N = 1200$ and
$P = 46~\kbt/8ab^{2}$; it formed SM2 as well. It also developed
a planar defect, this time owing to a
geometrical mismatch between the simulation box and the SM2 unit cell.
Note that for $a/b = 3$ and 4, SM2 formed even in simulations sampling
only rectangular box shapes. It is therefore more stable than
stretched fcc even when it cannot assume its equilibrium shape.

By contrast, ellipsoids with $a/b = 2$ and the oblate $a/b =
\frac{1}{3}$ showed no tendency to leave the initial
stretched-fcc structure. We studied each of these systems with eight
independent simulations. In none of them two preferred directions of
ellipsoid orientation developed. All of them melted to the
nematic phase on expansion, directly from stretched fcc.
But note that
the apparent stability of fcc in our simulations may well be due to a free
energy barrier, rather than indicating genuine stability.

\begin{figure}
\begin{center}
\includegraphics[ 
  width=\figwidth
  ]{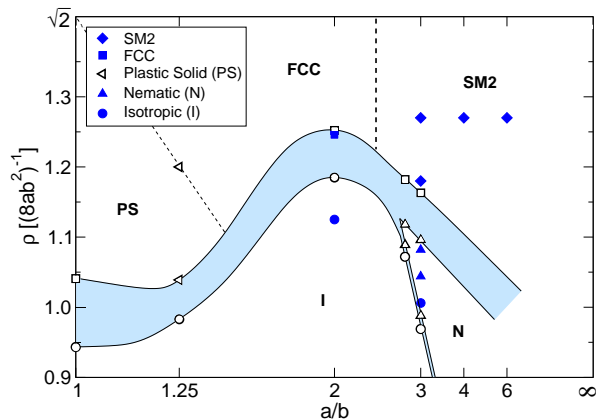}
\caption{\label{phasediagram} Phase diagram of hard, uniaxial
  ellipsoids, showing the results of Frenkel and Mulder (open symbols)
  \cite{frenkel.mulder:1985}, and their suggested phase boundaries and
  coexistence regions. The data points at $a/b = 1$ are taken from
  \cite{hoover.ree:1968}.
  We have inserted our state points (filled symbols) and a
  vertical dashed line to delimit the region in which we found SM2;
  but our data is insufficient to locate a phase boundary.
}
\end{center}
\end{figure}

In \Fig{phasediagram} we show a phase diagram of hard ellipsoids of
revolution. It includes part of the results of Frenkel and Mulder, and
their suggested phase boundaries and coexistence regions. We have
inserted our state points and a vertical dashed line to delimit the
region in which we found SM2; but our data is insufficient to locate
a phase boundary.

In the high density phase diagram of hard ellipsoids of revolution 
we have found a
crystal which is more stable than the stretched fcc structure proposed
by Frenkel and Mulder \cite{frenkel.mulder:1985}. The new phase, SM2,
has a simple monoclinic unit cell containing a basis of two
ellipsoids. The angle of inclination $\beta$ is a very soft degree of
freedom, whereas the other angles are not. At one value of $\beta$
($\approx 148.3^\circ$ for $a/b=3$),
close-packed SM2 is an instance of the family of packings introduced
by Donev et al.\ \cite{donev.etal:2004}. As for thermodynamic
stability, our results 
unequivocally remove the stretched fcc structure for aspect ratio $a/b
= 3$ from the phase diagram of hard, uniaxial ellipsoids. Our state
points for $a/b = 4$ and 6 suggest that this holds for the entire
range of $3 \leq a/b \leq 6$, and possibly beyond. However,
this does not prove that SM2 is the ground state. A procedure has been
developed for making almost 
monodisperse ellipsoids \cite{keville.etal:1991,ho.etal:1993}, which
are of colloidal size. Their behavior at water-air interfaces has been
studied \cite{basavaraj.etal:2006,loudet.yodh.pouligny:2006}; also 3D
structural properties of a sedimentation of these particles have been
successfully characterized \cite{mohraz.solomon:2005}. It would be
interesting to perform experiments probing colloidal crystals of ellipsoids.

This work was supported by the Emmy Noether Program and SFB TR6 of the
Deutsche Forschungsgemeinschaft (DFG), and the European Network of Excellence 
SoftComp. We are grateful
to the NIC J\"ulich for computing time, and to K.~Binder,
D.~Frenkel, M.~P.~Allen, J.~Vermant, A.~Donev, and
W.~A.~Siebel for helpful suggestions. 

\bibstyle{revtex}

\end{document}